# 마이크로스케일 음향흐름유동의 전산가시화
## Numerical Simulation of microscale acoustic streaming flow

*샤이야드 아티프 이크라르 [1,†], 박진수 [2,†], 무함마드 아프잘 [1], #성형진 [1]

*S. A. Iqrar [1,†], J. Park [2,†], M. Afzal [1], #H. J. Sung(hjsung@kaist.ac.kr)[1]

[1] 한국과학기술원 기계공학과, [2] 전남대학교 기계공학부

†These authors equally contributed to this work.

Key words : acoustofluidics, acoustic streaming flow, numerical simulation

## 1. Introduction

The efficient mixing of fluid samples in miniaturized total analysis systems is essential for numerous applications[1] including biological screening assays,[2] chemical extraction,[3] polymerization,[4] cell analysis and protein folding.[5] Miniaturized microfluidic platforms have recently emerged for microscale fluid mixing with their capabilities of high-throughput sample processing and reduced sample depletion.[6] However, microscale fluid mixing is inherently hampered due to the characteristics of low Reynolds number flows in diminutive channels. Microscale fluid mixing mainly depends on molecular diffusion and thus requires long processing time. In order to address the conundrum, a variety of active microfluidic approaches for swift and efficient mixing have been developed using electro-kinetic flow,[7-8] laser-induced flow,[9] magnetic stirring,[10] and acoustic streaming flow (ASF).[11-12] Among these techniques with external force fields, acoustofluidic fluid mixing has been acclaimed due to its on-demand, controllable, non-invasive and biocompatible nature. In this study, we performed numerical simulations of acoustic streaming flows, induced by with a highly localized surface acoustic waves (SAWs), in a microscale fluidic channel based on an acoustic wave attenuation model.

## 2. Methods

For numerical simulations of 3D microscale flows interacted with acoustic waves, COMSOL Multiphysics was utilized to model the effected fluid streamlines in the proximity of focused SAWs. The non-linear time-averaged body force and SAW attenuation model for the substrate/fluid interface and fluid domain are also considered. The attenuation lengths were analytically determined by using parametric relations $\alpha^{-1} \approx 12.8\lambda_{SAW}$ and $\beta^{-1} \approx 3e^6\lambda_{SAW}^2$ for the substrate/fluid interface and within the fluid itself, respectively[13] where $\alpha^{-1}$ is the coefficient of attenuation in substrate, $\lambda_{SAW}$ is wavelength of applied surface acoustic wave and $\beta^{-1}$ is coefficient of attenuation in fluid media. The effect of high-frequency SAWs originated from a focused interdigital transducer (FIDT) on the flow streamlines was numerically calculated in our numerical model. The grid independence was also confirmed by eliminating the influence of mesh parameters on the simulated acoustic velocities and streaming velocities in the laminar flow frequency domain to avoid any numerical error arising from grid convergence.

The COMSOL laminar flow model started with the definition of constant and variable parameters in both local and global domains. The significant constant parameters were defined globally for all physics while simulating the problem through a fully coupled scheme. The rest of the variables were manipulated in the extended definition step of each module separately to make it accessible for the respective model.

Analytical expressions of the attenuation coefficients for the substrate/fluid and fluid medium are given by

$$\alpha = \frac{\rho_f c_f}{\rho_s c_s \lambda} \quad , \quad \beta = \frac{b\omega^2}{\rho_f c_f^3}$$

Where $\rho$, $c$, $\lambda$, and $\omega$ indicate density, speed of sound, wavelength, and angular frequency, respectively (subscript $f$ and $s$ indicate fluid and substrate). The parameter $b = 4/3\mu + \mu'$ is a function of the dynamic and bulk viscosities. The above relations predict the length of an attenuated SAWs in the fluid/substrate domain, which has a direct relation with the frequency of applied waves. In second equation, the coefficient $\beta$ provides depreciated distance of leaky waves in a fluid, where the attenuation is the square of the applied acoustic wave frequency.

For domain definition, the 3D domain was built by using a geometric tool with a dimension of 550 × 80 × 650 in width, height and length of the microchannel. The width between the unsymmetrical vortices $L_w$ was approximately the width of the acoustic beam, which scaled with the aperture of a FIDT. The body force was calculated in the volumetric domain using the given relation

$$F_B = \rho\beta U(y,z)^2$$

Where $U$ is the first order fluid displacement velocity magnitude in the orthogonal direction to the fluid flow. Also, the velocity magnitude in the fluid was analytically determined by $U = \zeta\omega$, where $\zeta$ represents the overall attenuation in the substrate and fluid up to the first-order term while the rest of higher-order terms with reflection coefficient $R_i$ were terminated to avert complexity where $R_i = (\frac{Z_{i+1}-Z_i}{Z_{i+1}+Z_i})$. The $Z_{i+1}$ and $Z_i$ in the above relation are acoustic impedance values of the fluid/microchannel ceiling and fluid/substrate interfaces, and the acoustic impedance was calculated $Z_i = \rho_i c_i$. The analytical expression of the overall attenuation coefficient can be given as

$$\xi(y,z) = \xi_0(f_0(y,z)) + f_1(y,z)R_1 + f_2(y,z)R_1R_2 + ..$$

where $f(y,z)$ is the attenuation function that includes both the attenuation in substrate and fluid as well as the angle of refraction when a surface acoustic wave enters in a fluid at a certain angle which is $\theta_R = \sin^{-1}(c_f/c_s)$. The $\xi_0$ is initial displacement of SAWs, which was experimentally determined by laser Doppler velocimetry and particles image velocimetry. The first-order reduced attenuation function can be given as

$$\xi(y,z) = \xi_0(f_0(y,z)),$$

$$\xi(y,z) = \xi_0 \, (e^{-\alpha(y-z\tan\theta_R)} e^{-\beta z \sec\theta_R}).$$

The laminar flow physics was adopted to include fluid properties, Dirichlet boundary conditions for inlet velocity and outlet pressure, no-slip boundary condition, volumetric forces and incompressibility of fluid. Along with fluid flows, the trajectories of submicron particles under strong acoustic streaming effect was also simulated by using the particle tracing module.

### 3. Results and discussion

The magnitude of the time-averaged body force was directly related to the square of the acoustic velocity and attenuation coefficients. The numerical results show effective streaming in the fluid regime as compared to the fluid/substrate portion because the attenuation coefficient $\beta^{-1}$ in fluid is the square of the frequency of the SAWs while it only correlates with unit power of frequency in the substrate.

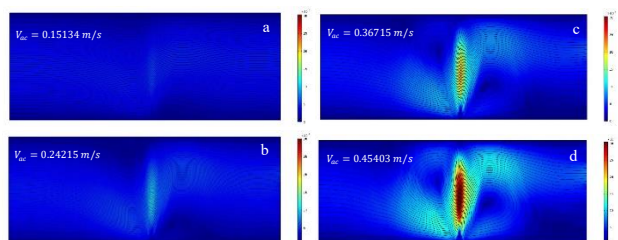

**Fig.1** Flow patterns in the vicinity of high frequency travelling surface acoustic waves placed orthogonal to the fluid flow direction. $v^* = v_{streaming}/v_{in}$, where value of $v^*$ increases the streaming velocity dominates over inlet velocity of fluid, thus streaming effect becomes significant for efficacious mixing of sample in microchannel. (a) Streaming effect is not prominent as $v^*$=1.99 (b) Gradual increase in acoustic velocity results in minor diversion of inlet streamlines where $v^*$=3.73 (c) and (d) represents the strong streaming effect where $v^*$= 9.35 and 11.83 respectively.

As shown in Fig. 1, the disturbance in flow streamlines was clearly observed as the ASFs become dominant over the lateral flow velocity. The simulated streamlines were plotted for increasing body force within the proximity of 11.4 $\mu m$ wavelength and 129.5 MHz acoustic waves acting orthogonal to a 3 mm/s (averaged) lateral flow in 550 $\mu m$ wide microchannel. A dimensionless parameter $v^*$ was introduce to represent the ratio of the time-averaged second-order streaming velocity ($v_2$) to the lateral inflow velocity ($v_f$). This specific parameter indicates the effectiveness of acoustic streaming in the fluid regime with applied SAWs. The increasing body force and streaming in the microchannel resulted in increasing value of $v^*$ in comparison with constant inflow velocity due to gradually elevated input power or acoustic wave amplitude as $v^* = v_2^{max}/v_f$. In Fig. 1(a), the effects of ASFs in the fluid streamlines was insignificant due to the small substrate velocity ($v^*$ = 1.99) of the orthogonally applied acoustic beam. In Fig. 1(b-d), in contrast, the prominent ASF-induced vortices were gradually developed due to the strong body force with increasing $v^*$. We found that the deflection of the flow streamlines due to the ASF-induced microscale vortices and resultant flow mixing were directly dependent on the amplitude of the acoustic field. It was also confirmed that the wave attenuation in the fluid as well as in the substrate, which in turn determined the size of the microscale vortices, was strongly correlated with the acoustic wavelength.

### 4. Conclusion

In the present study, we conducted numerical simulations of ASF-induced flow mixing by highly localized SAWs. We observed efficient ASFs in the microchannel to achieve efficient on-chip flow mixing in the microchannel. The results of numerical simulation advocate the effective mixing at microscale through acoustic attenuation of high-frequency SAWs generated from a FIDT.

### Acknowledgement

This work was supported by a National Research Foundation of Korea (NRF) grant (No. 2019022966) and the KUSTAR-KAIST Institute.